# The Hilbert transform of horizontal gaze position during natural image classification by saccades

Roberts Paeglis[1]*, Ivars Lacis[1], Anda Podniece[1], Nikolajs Sjakste[2]

[1]*Dept. of Vision Science, University of Latvia, 19 Raina Blvd., LV1036, Riga, Latvia*

[2]*Faculty of Medicine, University of Latvia, 1a Sarlotes Str., LV1001, Riga, Latvia*

Dated: October 4, 2006

**Abstract**

Eye movements are a behavioral response that can be involved in tasks as complicated as natural image classification. This report confirms that pro- and anti-saccades can be used by a volunteer to designate target (animal) or non-target images that were centered 16 degrees off the fixation point. With more than 86% correct responses, 11 participants responded to targets in 470 milliseconds on average, starting as quick as 245 milliseconds. Furthermore, tracking the gaze position is considered a powerful method in the studies of recognition as the saccade response times, ocular dynamics and the events around the response time can be calculated from the data sampled 240 times per second. The Hilbert transform is applied to obtain the analytic signal from the horizontal gaze position. Its amplitude and phase are used to describe differences between saccades that may testify to the recognition process.

## 1. Introduction

Two distinct lines of research are unified in the natural image classification. First, the vision researchers thus explore the nature and limits of the human visual system. It is done under the conditions of short image exposition and the demand for a rapid behavioral response (VanRullen, Thorpe, 2001), increased number of simultaneously presented items (Rousselet, Thorpe, Fabre-Thorpe, 2004), deteriorated image contrast (Macé, Thorpe, Fabre-Thorpe, 2005) etc. What types of pictures correspond best to the real world objects in human and animal visual system has also been a matter of detailed study (Bovet, Vauclair, 2000). It stirs the question of what higher cognitive processes are involved (VanRullen, Koch, Perona, 2002; Evans, Treisman, 2005). Second, the machine vision experts extract image features for the automated image classification. Grouping images by the color pixel distribution in a photograph has been shown to be a successful approach to categorize digital images (Greenspan, Goldberger, Ridel, 2001; Goldberger, Gordon, Greenspan, 2006). This is aimed at the image database search, for instance, data mining (Greenspan, Goldberger, Eshtet, 2001) or the medical catalogues (Lehmann et al, 2005). Image features that are used in the machine vision may not always be used to describe the human vision, as image metrics of the projections in the visual cortex are different from the metrics used to physically describe an image (Bertulis, Bulatov, 2001). On the other hand, understanding of the human vision can spill over to automated systems.

*Roberts.Paeglis@lu.lv

A typical response that a volunteer is asked to make to classify images is a button release (Rousslet, Thorpe, Fabre-Thorpe, 2004). The subjective response follows as soon as 250 ms after the image onset. In the objective measurements, differential neuromagnetic signal for the first pass starts after 150 ms for targets and distractors (VanRullen, Thorpe, 2001), resulting in behavioral response with a high precision of or above 80 per cent. This has led to alternative models for the neuronal communication in the visual pathway (Van Rullen, Gautrais, Delorme, Thorpe, 1998), which has been supported by the data from other sensory systems (Johansson, Birznieks, 2004; VanRullen, Guyonneau, Thorpe, 2005). Research suggests that the differential neural activity, which is evoked by semantic categorization of objects into groups like natural objects (animals, plants) or man-made objects (furniture, clothing), is not task dependent (Löw et al., 2003). If one compares the time for cortical processes for isolated objects as detected by Löw et al., and that for objects in context (VanRullen, Koch, Perona, 2002; VanRullen, Thorpe, 2001), it can be argued that the context facilitates classification of object images but is not the decisive factor.

In a recent paper, Kirchner and Thorpe (Kirchner, Thorpe, 2006) report that eye movements may emerge as a faster way to respond in a classification task. With the minimum saccade response time of 120 milliseconds, this is faster than the differences in the ERP (Thorpe, Fize, Marlot, 1996) and has inspired possibilities that each pass actually uses a different neuronal path. Eye movements are a fast and precise corporal response under voluntary control. The benefit of using saccades in the image classification task is manifold. Besides keeping the subjective response within the oculo-motor system, it equips the researcher with supplementary data to analyze in addition to response time. The gaze trajectories, eye movement dynamics and the site of destination are but other facets of the classification process.

Saccade control involves an overlap between top-down and bottom-up processing that is driven by visual information (Riesenhuber, Poggio, 2000; Mosimann, Felblinger, Colloby, Müri, 2004). We have set a pro- and anti-saccade task to classify digital images where the number of images shown corresponds to classical trends in saccadic research (Mosimann, Felblinger, Colloby, Müri, 2004; Leigh, Kennard, 2004), whereas division into categories is taken from the mainstream research (Rousslet, Thorpe, Fabre-Thorpe, 2004). This has permitted us both to calculate statistics for the response times in categorization and to review individual performance case by case. In addition, research volunteers were interviewed for their subjective perception after the experiments.

2. Methods

*2.1. Participants*

Eleven adult volunteers (21 to 28 years old) had their eye movements recorded with the aid of the IR corneal reflection. Their vision was normal by convention (Visus of 0.9 or above) or corrected to normal. They had varied exposure and knowledge about the images of animals as they had different educational backgrounds, ranging from life sciences to humanities. Every individual participated with an informed consent. The participants were made aware of the fact that the experimental procedures would be interrupted in the case of their ocular or other discomfort.

*2.2. Experimental set-up*

Before the experiment, participants had adapted to the level of screen illumination in a dim room. The color images were presented on a gray background of a 17 inch LCD screen (1280 by 1024 pixels) with the vertical refresh frequency of 75 Hz and pixel reaction time of 8 milliseconds. The volunteers were seated 40 cm from the screen. The mean intensity of target and non-target images was matched.

Eye movements were recorded by the *iViewX Hi-Speed 240* IR device (SMI, Germany). The data of the gaze position were saved to experimenter's PC at the rate of 240 Hz. The image presentation of the participant's PC and data recording on the experimenter's PC was proved to be synchronous within 1 ms.

The images were shown on the side of the eye that was tracked by the IR camera. For each eye being recorded, the participant was presented 30 images in the training set and 30 images in the experimental set (60 images per eye). In each set, ten images in a random sequence were targets that by instructions required a pro-saccade, while the correct response to a non-target was an anti-saccade. While instructing the volunteers, the rapid response was stressed by urging them to make an eye movement while the image was still on the screen. Images were presented on the PC screen for 300 milliseconds. The instructions were repeated after the training set (before the experiment *per se*). Images were resized from a commercially available collection, targets being animals ranging from invertebrates (spiders, snails) to vertebrates (fish, birds, reptiles, mammals). Half of the non-target images were natural landscapes the other half was man made objects like versatile means of transport.

*2.3. Software implementation*

The images were shown at random by a *Visual Basic* code that was checked to perform the operations in the range of microseconds on a Pentium 4 PC. The images stretched 300 by 225 pixels or 8.1 deg wide on the screen. They were centered 16 deg off the center of the screen (the fixational point), so as to make it possible to move the images in either direction on the screen or resize them should it be necessary for a further research. After an image has been shown, the participants had a random period of 4000 to 6000 milliseconds (4 to 6 sec) to re-fixate the white central point that was constantly on the screen. At the same time, two smaller dots were shown on the sites of pro- and anti-saccade. The proper fixation was later approved in the eye movement data.

For five participants 30 training and 30 experimental images were presented. Six individuals had the same procedure repeated for the other eye (4x30 images presented in total) to test possible dependence of the response time on the presentation side.

The beginning of the first saccade after image onset was calculated by the ILAB 3.6.4. (Gitelman, 2002) under MATLAB when artifacts (blinks) had been filtered out, and was confirmed by tracing the raw data for the gaze co-ordinates. The horizontal component of gaze position was equated to the "signal" to which signal processing, namely, the Hilbert transform was applied under MATLAB.

# 3. Results

## 3.1. Response times in the pro- and anti-saccade task

With different kinds of animals as target images, the pro-saccade response times of 450 ± 160 ms and the anti-saccade reaction times of 470 ± 140 ms for 11 individuals had been achieved. The first correct pro-saccade was initiated 245 to 603 milliseconds after the image onset, the correct anti-saccades started 348 to 652 milliseconds after the image has appeared. The standard deviations varied by individual from 84 to 265 milliseconds. On average, the digital images were correctly classified in 86% of cases, with one individual exposing consistent problem with suppression of the visual grasp reflex. Albeit this individual (NV) managed to make the first eye movement in less than 287 milliseconds and a correcting saccade within 600 milliseconds after image onset, the following eye movements were not considered to be correct responses and were discarded from the statistical analysis. In the same manner, all the following correcting saccades for other individuals were considered to be erroneous responses if the first saccade was misdirected, once the gaze has left 1 deg around the fixation point.

## 3.2. Performance and ocular dominance

For six subjects, the inter-ocular distance and the optical correction, if one was needed, permitted us to obtain reliable measurements of saccadic response for each eye. When t-tests were applied for the statistical significance, we did not find conclusive evidence for a faster saccadic response for the images presented on the side of the dominant eye. The statistical significance for the response times of both eyes for four individual ranged from p=0.06 to p=0.94 (mean p=0.29, T-test), while the fifth individual had p=0.014 and the sixth p<0.01. The analysis revealed that the statistical significance for the latter was due to faster reaction to target images (574 to 369 ms) in favor of the eye that was tracked later. In both cases, however, this was the non-dominant eye. Upon further investigation the later two individuals were found to have benefited the most from the trial set and the repeated instructions after it. In other words, they have successfully used 60 presentations for the first eye and the following 30 trial presentations to the second eye to make general inferences about the task and adapt to it. Thus for some but not all individuals the learning effect impacts the performance and expanding the image sets can lead to shorter response times provided that it is tolerated by the oculo-motor system.

*3.3. Continuous signal vs. single data point*

Eye movement data were collected 240 data points per second, in this way the researcher has the advantage of having complimentary data to the saccadic response time. In our task, the coordinates of the gaze position can be visualized as the "signal" like the laser beam or other intensity. At some point a voluntary saccade is launched and is followed by a brief fixation which is similar to a step or Heaviside function. The pro-saccades are all peaks and the anti-saccades are all wells (or vice versa, depending on the side of image presentation). This approach alleviates data processing as after the normalization one group has positive values and the other negative ones, to which the image sequence list can be compared.

Besides extracting the timing of saccadic response, it is instructive to contrast and compare the saccade dynamics, such as the velocity and acceleration, saccade size and the nature of the following correcting and return saccades for different individuals. We calculated the amplitudes and omitted micro-saccades from the analysis. In order to access the measurements, we plotted the velocities of proper saccades in the main sequence coordinates like is reported elsewhere (Harwood, Mezey, Harris, 1999).

If the main sequence analysis is viewed as the relation of the maximum velocity of a saccade over the average velocity of it, this relation basically represents the dynamics of eye acceleration. For two individuals with the high correct response rate no difference could be credibly stated for the leftward or rightward velocity of the person's eyes,

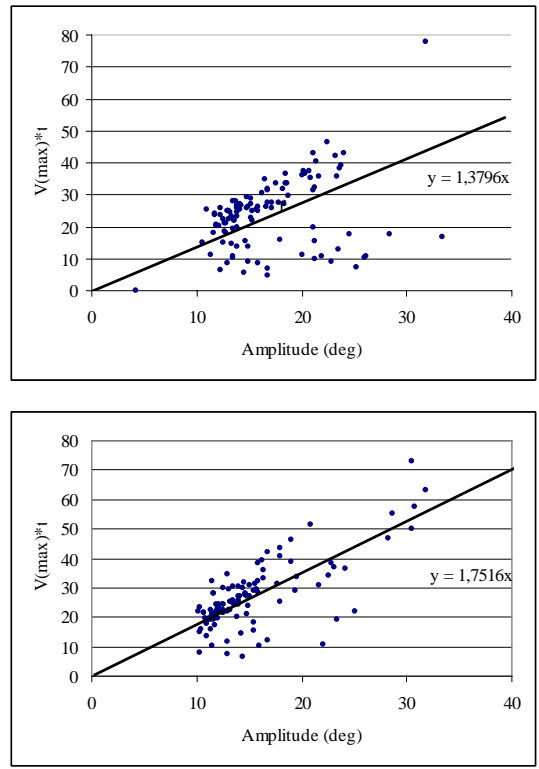

**Fig. 1.** Saccade maximum velocities plotted for the main sequence analysis.
Upper graph: person KB, lower: person EB.

and were plotted together for both eyes (Fig. 1).

Notwithstanding that, the dynamics of acceleration for both volunteers differed (as represented by the regression line coefficients, 1.38 vs 1.75), which may testify to discrepancies in the ocular biomechanics but possibly also to differences in innervations due to the natural image recognition process. Return saccades to the central fixation point are a controversial issue by themselves as for some but not all distracters the return has been accomplished in multiple steps rather than in one sweep.

*3. 4. Eye movement "fingerprints" from Hilbert transform of gaze coordinates*

Tracking the participant's eye, like brain imaging, opens new sources for information concerning the recognition process in time. It does so by amassing raw data, a trial of a 100 sets, 4.2 seconds each, produces more than $10^5$ data points for the horizontal gaze position alone. The ambition then is to find descriptors or "fingerprints" that allow us to explore and classify the saccadic resonses and compare them to the targets seen.

Most of what we see in the data are small scale changes around zero when the center of the screen is designated as the point (0,0). In some cases, there is some larger gaze instability before a saccade is launched (Fig. 2), which is followed by a large amplitude change. The fixation part and the entire set can be equally weakly approximated by several functions, such as the sum of harmonic functions or polynomials, and thus remains uninformative. However, the changes in the gaze position during a trial contain the information about the processes before and after the saccadic response, as well as the parameters of this response.

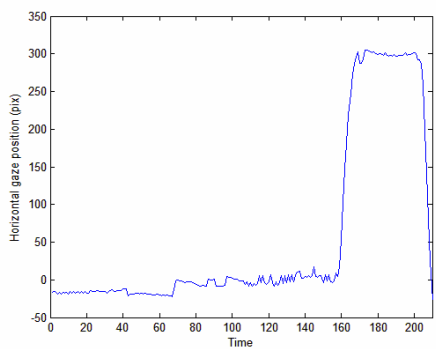

**Fig. 2.** Gaze position fluctuations are observed before a saccade is launched. One X-axis unit is 4.2 milliseconds.

The Hilbert transform, unlike many other mathematical methods that can be applied to frequently sampled data, returns the results of the same domain as the initial "signal" (data). Adding the imaginary unit times the Hilbert transform of data to the data itself is a means to transfigure the real "signal" into the complex analytic signal, whose amplitude and phase are informative of the "signal" details. The Hilbert transform reveals interference or interaction (DeShazer, Breban, Ott, Roy, 2001) that is not readily legible from the original data, the phase of this transform describes changes in the field envelope, i.e., the long run changes in the response such as the learning effects. Small amplitude fluctuations in the gaze fixation before a saccade is launched, as well as the saccades, lend themselves to such analysis (Fig. 3).

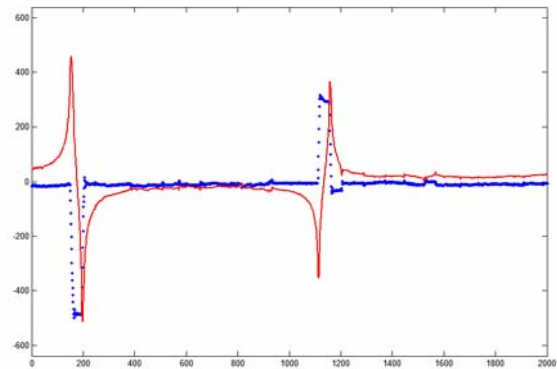

**Fig. 3.** The Hilbert transforms for the saccades to opposite horizontal directions. Dotted line: gaze position; solid line: the Hilbert transform. Horizontal axis: time (by 4.2 millisecond units), vertical axis: pixels.

We have calculated the Hilbert transform for 180 milliseconds since the trial start and divided the data into two groups: fixations before saccades to targets (animals) and before those to non-targets. This is a period when no saccade has been launched, so they did not influence the shape of the transform. To quantify the trend in responses before a saccadic response has been made, we pooled together the absolute values of the analytic signal. Starting approximately 120 milliseconds (27 points times 4.2 milliseconds) from the image presentation, the transformations of the horizontal gaze position at the central fixation cross zero and diverge (Fig. 4). We are fully aware of the fact that these pilot results may be specific to population and should not be taken as a reference. What we do propose is that the aforementioned mathematical construct is of use in exploring changes in fixation.

The Hilbert transform, if it is applied on the scale of the entire set of horizontal coordinates, becomes a filter for saccades as its peaks correspond to the saccade ends. (Fig. 5).

the eye movement. Its derivative assumes high value at some points during fixation also, especially after the movement is made and the gaze returns to refixate. To these ends, Fig. 6 captures horizontal gaze position during three image presentations. In the first one, a correct anti-saccade is made (downwards in the graph). At the next image, an incorrect saccade has been launched 350 milliseconds from image onset and it seems to be terminated before usual (6.4 deg), then in 630 milliseconds an attempt is made to correct the response with a saccade (upwards), which was not considered in statistics. The last peak is a correct saccade to the target's side. The gap between the erroneous anti-saccade and the correcting pro-saccade is notable for the rigidness of fixation (small values of the complex phase derivative), whereas the period right after the correcting movement is characterized by fluctuations that are large compared to other trials (see the first and the last spike in Fig. 6.

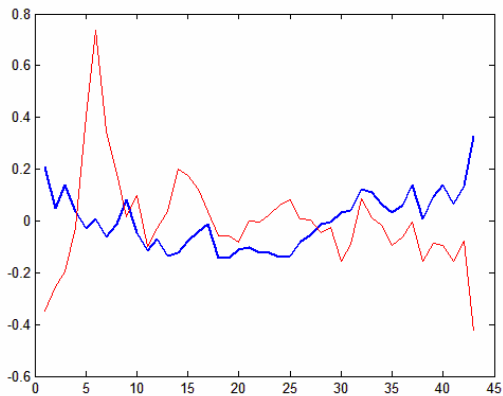

**Fig. 4.** The Hilbert transform of fixations after the image has appeared. Responses before a pro-saccade (think line) and an anti-saccade (see text). Horizontal axis: time (4.2 millisecond units), vertical axis: average value of the Hilbert transform.

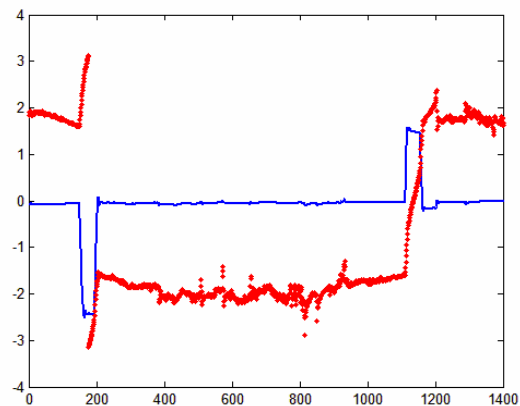

**Fig. 5.** After the gaze position (solid line, value divided by 200) is transformed into a complex analytic signal (dotted), its phase experiences jumps of $2\pi$ under conditions that can be modified.

The mathematically constructed complex phase, in turn, experiences a jump of $2\pi$ at one type of saccades and not the other (depending on which side is the task for the pro-saccades), and so it does irrespective of the numerical value (in pixels) of

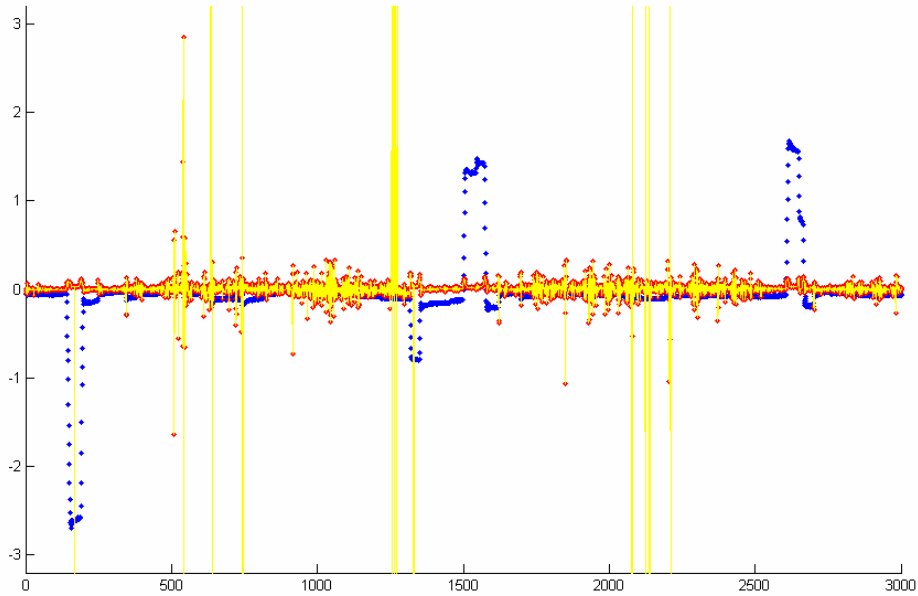

**Fig. 6.** If the horizontal gaze position (dotted, value dividend by 200 to fit the plot) is transformed into the analytic signal, the derivative of its complex phase experiences peaks under some conditions during fixations in the trials. Horizontal axis: time in 4.2 millisecond units, vertical axis: values

## 4. Discussion

Recent advancements in natural image recognition (Kirchner, Thorpe, 2006) have raised the questions concerning the correspondence of behavioral response times in the eye movement and button release data. Kirchner and Thorpe argue that while some images are faster classified by the eye movement paradigm, this does not hold for the other modes of response. The applicability of the eye movements to classify images has intrigued us to use pro- and anti-saccades to distinguish target and non-target images. If eye movements are tracked continuously during a classification task, several issues related to the moment of response and the surrounding milliseconds emerge.

### 4.1. Task-specific training

The volunteer's educational background did not appear to be of benefit in the image classification experiments. However, we chose a set of images in line with several other protocols of pro- and anti-saccade research (Mosimann, Felblinger, Colloby, Müri, 2004), which is restricted if compared to the reported data of Kirchner and Thorpe (Kirchner, Thorpe, 2006). None of the participants proved to perform faster than 245 milliseconds after 60 images have been seen. On one hand, we acknowledge that the statistical analysis we use differed from the reported data. On the other hand, the results combined can hint to the plasticity of the visual system. Even though completing saccades in several series of 80 images each is a visually demanding task, it can be learned and performed with high precision. This, however, brings into focus inter-personal differences in trainability and motivation, as we have seen rapid improvement of saccade response times in some but not all subjects. The research has also affirmed that a person can be taught to designate classification by making either a pro-saccade or an anti-saccade instead of a pro-saccade to one of two images. This at least partially explains slower response times than those reported (Rousslet, Thorpe, Fabre-Thorpe, 2004; Macé, Thorpe, Fabre-Thorpe, 2005), as saccade suppression must first happen. A forced choice of two alternative images presents the visual system a possibility to weigh and compare the "animal-likeness" of two images, where only partial representation within the visual system may be sufficient. With only one image presented, no mutual exclusiveness is engaged and a probabilistic decision must be made once sufficient processed information has accumulated.

The long saccadic response times and their large variation are not typical for a task demanding a pro-saccade to a target onset (like a saccade to a lit LED), but are more akin to a delayed saccade task which involves some cognitive access (Mosimann, Felblinger, Colloby, Müri, 2004).

A fraction of erroneous responses were in fact later corrected with another eye-movement, which highlights the concepts of the Signal Detection Theory. The volunteers that admitted to a high personal motivation to perform fast in a novel task at a succeeding interview were more tended to respond faster with an incorrect type of saccade. Thus they had set a low threshold with low miss rates but high number of false alarms. Alternatively, one could conclude that the visual grasp reflex is not always well suppressed in a task of this kind.

*4.2. Types of information from eye movements*

Eye tracking equips the researchers with a potential for additional information sources but it does not enable them to get response times in the real time from the raw data. Instead, many data points of gaze position and pupil size are grasped every second, in our case amounting to 240 points for each traced variable. Posterior analysis is needed to extract the saccade start and end times from the recordings.

The facility to visualize eye movement data for the entire trial is a means for screening saccades that reach the image location from any small amplitude eye movements. It is also a way to examine the quality of responses made. For the purposes of our study, all but the very first large size eye movements were discarded. From the other point of view, the common errors and their corrections that are grabbed in the data may appear explanatory for the recognition process if case-by-case studies are performed.

The saccades in this image classification task had been characterized by their dynamics, namely, the mean and maximum velocity and maximum acceleration. The different slopes of the main sequence that most saccades aligned to (Fig. 1) could be explained with physiological differences among the volunteers, however, velocities of some saccades formed a distinct cluster that was the most populated for the person KB. These spurious cases could not be fully accounted for by measurement errors, and we failed to find any prominent differences in other parameters of the suspect saccades.

Of less informative value we consider to be the end points were the gaze lands after a pro-saccade. It has been shown (van der Linde, Rajashaker, Cormack, Bovik, 2005) that non-random and salient image points are preferably selected for the first fixations from a relatively homogeneous scene, like a grass field. It is also argued in other reports how classification can rely on the amount of object features memorized (Peters, Gabbiani, Koch, 2003). Therefore tasks could be further designed where no markings are set before the images are flashed, either as an alternative forced choice or one image at a time. Even though saccade sizes vary in the obtained results (Fig. 1), just as the animal head and body positions in the images do, we had not observed any conclusive pattern in the variability of saccade sizes. This could be anticipated as the white circles shown before images apparently are strong attractors. The white cueing points during the central fixation create a bias that deserves further exploration.

One could in principle analyze the spots in the target images to which the saccades had been targeted, and look for them at the image saliency maps. However, this would ask for a modified experimental procedure as in our case the saccades in majority of cases were launched after the image has disappeared (after 300 milliseconds ) and so the memory-guided control should be borne in mind.

*4.3. The moments before and after the response*

The interest in eye movements to classify images may be extended to other features besides the saccade response times. The challenge the researcher faces, though, is that each "fixation and saccade" trial does not lend itself easily to geometrical analysis. Small amplitude (less than 2 degrees) movements around fixations are followed by saccades that exceed 20 degrees (Fig. 2). The validity of regression equations is therefore undermined. The phase of the Hilbert transform is a sort of normalization, as for different eye movement sizes its changes are held within $2\pi$.

We chose the horizontal gaze position as the variable that is task oriented and changes the most. The data have been checked for the quality, blinks being filtered out. Insufficient care of data quality can yield spurious results, as the mathematical transforms are sensitive to any interruptions in the data stream, like zeros during blinks or otherwise lost corneal reflections. Furthermore, the application of the Hilbert transform in the laser science is for the exploration of interference and interactions (DeShazer, Breban, Ott, Roy, 2001). If any feedback can be suspected in the oculo-motor system in the case of multiple stages of image processing, difference of a certain period of time (in milliseconds) should be probed. This may eventually shed some light to the debate of whether saccades are ballistic movements or they can be corrected on flight. The erroneous saccade in Fig. 6 had a smaller size, it is then worth clarifying at what stage the incorrect response or the visual grasp reflex can be halted.

The Hilbert transform, the amplitude and the phase of the analytic signal have hysteresis in the sense that minor changes at a point influence the appearance of the transform for the several surrounding points. Since phase jumps can occur at different amplitudes, this may become a drawback, since noisy fixations and multiple stage responses cause distinct phase jumps, unless one defines the physiologically plausible elements of the signal to probe with this approach.

It could be further inquired into the nature of fixations before and after a correcting saccade is made, as in Fig. 6. Do the fluctuations in the fixating convey the underlying neural processes? If it is only noise, what influences the changes in the magnitude of this noise as seen from the Hilbert transform?

We hypothesize that the fluctuations in the complex phase between saccades describe preparatory processes during recognition and before the response (Fig. 6). After a saccade is made, the gaze is stabilized and can testify to some movement inhibition that is necessary for the visual process.

**Conclusions**

We consider that the possibility of frequently sampled eye tracking is the main power of the eye movement responses in image classification tasks. In addition to response times, the ocular dynamics, the nature of errors and events around the response can be studied when the Hilbert transform is applied. The reported tasks have supported the value of pro- and anti-saccade tasks in telling animal and non-animal images apart.

**Acknowledgements**
This research is supported by the European Social Fund and the University of Latvia.
The authors are grateful to Vyacheslavs Kashcheyevs for apt Physics comments.

**References**

Bovet D., Vauclair J. (2000) Picture recognition n animals and humans, Behavioural Brain Research, 109, pp. 143–163.

Bertulis A., Bulatov A. (2001) Distorsions of Length Perception in Human Vision. Biomedicine, 1 (1), Jan.–Jun., pp. 3–26.

DeShazer D., Breban R., Ott E., Roy R. (2001) Detecting Phase Synchronization in a Chaotic Laser Array, Physics Review Letters, 87 (4), 23 July, pp. 0044101-1-0044101-2.

Evans K. K., Treisman A. (2005) Perception of Objects in Natural Scenes Is It Really Attention Free? Journal of Experimental Psychology, Human Perception and Performance, 31 (6), pp. 1476–1492.

Gitelman D. R. (2002) ILAB: A program for postexperimental eye movement analysis, Behavior Research Methods, Instruments, & Computers, 34 (4), pp. 605–612.

Goldberger J., Gordon Sh, Greenspan H. (2006) Unsupervised image-set clustering using an information theoretic framework, IEEE Trans. on Image Processing, 15(2), pp. 449–458.

Greenspan H., Goldberger J., Eshet I. (2001) Mixture model for face – color modeling and segmentation. Pattern Recognitions Letters, 22, pp. 1525–1536.

Greenspan H., Goldberger J., Ridel L. (2001) A continuous probabilistic framework for image representation and matching. Journal of Computer Vision an Image Understanding, Dec., 84, pp 384–406.

Harwood M. R., Mezey L. E., Harris C. M. (1999) The Spectral Main Sequence of Human Saccades, Journal of Neuroscience, 19 (20), October 15, pp. 9098–9106.

Johansson R. S., Birznieks I. (2004) First spikes in ensembles of human tactile afferents code complex spatial fingertip events, Nature Neurosci., 7 (2), February, pp. 170–177.

Kirchner H., Thorpe S. (2006) Ultra-rapid object detection with saccadic eye movements: Visual processing speed revisited, Vis. Res., 46, pp. 1762–1776.

Lehmann T. M., Güld M. O., Deselaers T., Keysers D., Schubert H, Spitzer K., Ney H., Wein B. B. (2005) Automatic categorization of medical images for content-based retrieval and data mining, Comp. Med. Imag., 29, pp. 143–155.

Leigh R. J., Kennard C. (2004) Using saccades as a research tool in the clinical neurosciences, Brain, 127, pp. 460–477.

Löw A., Bentin Sh., Rockstroh B., Silberman Y., Gomolla A., Cohen R., Elbert T. (2003) Semantic Categorization in the Human Brain: Spatiotemporal Dynamics Revealed by Magnetoencephalography, Psychoogical Science, 14 (4), July, pp. 367–372.

Macé M. J.-M., Thorpe S. J., Fabre-Thorpe M. (2005) Rapid categorization of achromatic natural scenes: how robust at very low contrasts? European Journal of Neuroscience, 21, pp. 2007–2018.

Mosimann U. P., Felblinger J., Colloby S. J., Müri R. M. (2004) Verbal instructions and top-down saccade control, Exp. Brain Res., 159, pp. 263–267.

Peters R. J., Gabbiani F., Koch C. (2003) Human visual object categorization can be described by models with low memory capacity. Vis. Res., 43, pp. 2265–2280.

Riesenhuber M., Poggio T. (2000) Models of object recognition, Nature Neurosci. Supplement, 3, November, pp. 1199–1204.

Rousslet G.A., Thorpe S.J., Fabre-Thorpe M. (2004) Processing of one, two our four natural scenes in humans: the limits of paralelism, Vis. Res., 44, pp. 877–894.

Thorpe, S., Fize, D., and Marlot, C. (1996) Speed of processing in the human visual system. Nature, 381, pp. 520–522.

van der Linde I., Rajashaker U., Cormack L. K., Bovik A. C. (2005) A study of human recognition rates for foveola-sized image patches selected from initial and final fixations on calibrated natural images, SPIE Human Vision & Electronic Imaging X, 5666, March, pp. 27–38.

Van Rullen R., Gautrais J., Delorme A., Thorpe S. (1998) Face processing using one spike per neurone, BioSystems, 48, pp. 229–239.

VanRullen R., Guyonneau R., Thorpe S. J. (2005) Spike times make sense, TRENDS in Neuroscience, 28 (1), pp. 1–4.

VanRullen R., Koch Ch., Perona P. (2002) Rapid natural scene categorization in the near absence of attention, PNAS, 99., pp. 9596–9601.

VanRullen R., Thorpe S. J. (2001) Is it a bird? Is it a plane? Ultra – rapid visual categorization of natural and artifactual objects. Perception, 30, pp. 655–668.